\documentstyle[aps,prl,cite,enumerate,verbatim]{revtex}
\begin{document}
\title{Clauser-Horne inequality for qutrits}
\author{Dagomir Kaszlikowski,$^{1,2}$, L. C. Kwek,$^{1,4}$, Jing-Ling Chen,$^{1,3}$   
Marek \.Zukowski,$^{5}$ and C. H. Oh$^1$}
\address{$^1$Department of Physics, National University of Singapore, 10
Kent Ridge Crescent, Singapore 119260 \\
$^2$Instytut Fizyki Do\'swiadczalnej, Uniwersytet Gda\'nski, PL-80-952, Gda\'nsk, Poland,\\
$^3$ Laboratory of Computational Physics, 
Institute of Applied Physics and Computational Mathematics, \\
P.O. Box 8009(26), Beijing 100088, People's Republic of China \\
$^4$National Institute of Education, Nanyang Technological University, 1 Nanyang Walk, Singapore 639798\\
$^5$Instytut Fizyki
Teoretycznej i Astrofizyki, Uniwersytet Gda\'nski, PL-80-952,
Gda\'nsk, Poland.}

\maketitle

\begin{abstract}
In this brief report we show the new Bell-Clauser-Horne inequality for
two entangled three dimensional quantum systems (so called qutrits).  
This inequality
is violated by a maximally entangled state of two qutrits
observed via symmetric three input and three output port
beamsplitter only if the amount of noise in the system equals ${11-6\sqrt3 \over 2}\approx 0.308$. This
is in perfect agreement with the previous numerical calculations presented in Kaszlikowski
{\it et. al.} Phys. Rev. Lett. {\bf 85}, 4418 (2000).  
\end{abstract}

\section{Introduction}

It is well known that the sufficient and necessary condition for the lack of existence
of a local realistic description of two entangled qubits in an experiment
in which Alice and Bob measure two dichotomic observables is the violation of
Clauser-Horne (CH) \cite{CH,FINE} inequality (in fact, there are four CH inequalities and
at least one of them must be violated). However, for a system of two entangled
three dimensional quantum objects,there has hitherto been no such inequality. Numerical
calculations based on the method of linear programming \cite{KASZLIKOWSKI,DURT} as well as
a subsequent analytical proof \cite{CHEN} clearly demonstrate that violation of 
local realism for qutrits is stronger
than for qubits. However, numerical calculations do not allow us to 
fully understand the nature of quantum
correlations\footnote{The analytical proof \cite{CHEN} of 
the results for qutrits obtained in \cite{KASZLIKOWSKI} cannot be applied to an arbitrary quantum state
and arbitrary observables.} and the possibility of their local and realistic description. 
In short, we need a Bell inequality.

At this junction, it is worth recalling that for two entangled qubits we have in fact two 
inequalities. The Clauser-Horne-Shimony-Holt (CHSH) inequality \cite{CHSH} imposes 
constraints on the properly defined 
{\it correlation} functions that can be measured in an experiment. 
However, this inequality provides only
necessary condition for a local realistic description of the experiment (it only gives necessary
and sufficient conditions for the existence of local and realistic description of the correlation
function). 
A second inequality is the CH inequality \cite{CH} which imposes constraints 
on the measured {\it probabilities}. It has been proved
by Fine \cite{FINE} that this inequality (more precisely four inequalities, see the remark above) 
is a necessary and sufficient condition for 
the existence of local hidden variables. Indeed, CH inequality is more fundamental 
than CHSH one. It stems from the fact that the fundamental notion in any Bell experiment 
are actually probabilities and not the correlation function. Moreover, it is not possible 
to define correlation function in some fixed manner.

In this short report, we present a CH-type inequality which 
relies on the probabilities measured in
a Bell experiment with two entangled qutrits (the 
inequality for the some well-defined correlation function
will be presented in a separated paper). We do not know if this inequality gives the 
sufficient condition for the existence of local realism in this case. However, 
the violation of local realism predicted by 
this inequality is found to be in a perfect agreement with the results obtained in \cite{KASZLIKOWSKI, CHEN}.
Moreover, the previous results provides {\it necessary and sufficient} conditions for local realism. Thus, 
it strongly suggests that we have the sufficiency condition in our case.

\section{Inequality}

In a Bell type experiment with two
qutrits Alice and Bob measure two trichotomic (there are three possible outcomes) 
observables: $A_1,A_2$ for Alice and $B_1,B_2$ for Bob. 
The outcomes of the measurement of observable
$A_k$ ($k=1,2$) at
Alice's side we denote by $a_k=1,2,3$ whereas the outcomes of the measurement of observable
$B_l$ ($l=1,2$) at
Bob's side we denote by $b_l=1,2,3$.
For each pair of observables $A_k, B_l$ 
($k,l=1,2$) we calculate the joint quantum probabilities $P_{QM}^{kl}(a_k;b_l)$, i.e., the 
probabilities of obtaining by Alice and Bob simultaneously the results $a_k$ and $b_l$ (coincidence 
"clicks" of detectors) and single quantum probabilities $P_{QM}^{k}(a_k),Q_{QM}^{l}(b_l)$, i.e., 
the probabilities of obtaining the result $a_k$ by Alice irrespective of Bob's outcome and result $b_k$ by Bob
irrespective of Alice's result.

A local realistic description assumes the existence of a joint probability 
distribution $P_{LR}(a_1,a_2;b_1,b_2)$, in which one can recover quantum probabilities 
from the 
marginal probabilities, i.e.,
\begin{eqnarray}
&&P_{LR}^{kl}(a_k;b_l)=\sum_{a_{k+1}=1}^{3}\sum_{b_{l+1}=1}^{3}P_{LR}(a_1,a_2;b_1,b_2)\nonumber\\
&&P_{LR}^{k}(a_k)=\sum_{a_{k+1}=1}^{3}\sum_{b_1=1}^3\sum_{b_2=1}^3P_{LR}(a_1,a_2;b_1,b_2)\nonumber\\
&&Q_{LR}^{l}(b_l)=\sum_{a_1=1}^{3}\sum_{a_2=1}^{3}\sum_{b_{l+1}=1}^{3}P_{LR}(a_1,a_2;b_1,b_2),
\label{joint}
\end{eqnarray} 
where $k+1$ and $l+1$ are modulo 2. In other words, if there exists
a local realistic description of an experiment, one 
has $P_{QM}^{kl}(a_k;b_l)=P_{LR}^{kl}(a_k;b_l)$ etc.
With these formulae (\ref{joint}) one can prove that  
\begin{eqnarray}
&&P_{LR}^{11}(2;1)+P_{LR}^{12}(2;1)-P_{LR}^{21}(2;1)+P_{LR}^{22}(2;1)+\nonumber\\
&&P_{LR}^{11}(1;2)+
P_{LR}^{12}(1;2)-P_{LR}^{21}(1;2)+P_{LR}^{22}(1;2)+\nonumber\\
&&P_{LR}^{11}(2;2)+P_{LR}^{12}(1;1)-P_{LR}^{21}(2;2)+P_{LR}^{22}(2;2)-\nonumber\\
&&P_{LR}^{1}(1)-P_{LR}^{1}(2)-Q_{LR}^{2}(1)-Q_{LR}^{2}(2)\leq 0.
\label{inequality}
\end{eqnarray}
The proof is straightforward but laborious and it is given in the Appendix. 
The above inequality is the necessary condition for
the existence of a local and realistic description of the considered experiment.

\section{Violation}

We now show that the above inequality is violated by quantum mechanics. To this end, let
us consider the following Bell experiment. The source produces maximally entangled state $|\psi\rangle$
of two qutrits

\begin{eqnarray}
&&|\psi\rangle = {1\over\sqrt 3}(|1\rangle_A |1\rangle_B+
|2\rangle_A |2\rangle_B+|3\rangle_A |3\rangle_B)
\label{state}
\end{eqnarray}
where $|k\rangle_{A}$ and $|k\rangle_{B}$ describe $k$-th basis state of
the qutrit $A$ and $B$ respectively. Such a state can be prepared
with pairs of photons using parametric down conversion(see
\cite{TRITTERS}), in which case kets $|k\rangle_{A}$ and $|k\rangle_{B}$
denotes photons propagating to Alice and Bob in mode $k$. Starting with this
state, Alice and Bob measure two trichotomic observables defined
by 6-port (three input and three output ports) beam splitter.
The extended theory of
such devices can be found in \cite{TRITTERS}. 
A brief description is provided below.

{\it Unbiased} $6$-port beamsplitter, which is called tritter \cite{TRITTERS},
is a device with the following property: if a photon enters
any single input port (out of the $3$), there is equal likelihood that
it leaves one of the three output states. 
In fact, one can always construct a tritter with the distinguishing trait 
that the elements of its unitary transition matrix, $\hat{T}$, are
{\it solely} powers of the complex number,$\alpha=\exp{(i2\pi/3)},$
namely $T_{kl}= \frac{1}{\sqrt{3}}\alpha^{(k-1)(l-1)}.$
Indeed, we can put a phase shifter in front of $i$-th input 
port of the tritter, changing the phase of the incoming photon by $\phi_i$. 
The three phase
shifts, which we denote for convenience as a ``vector" of phase shifts 
$\vec{\phi}=(\phi_1,\phi_2,\phi_3)$, are macroscopic local parameters
that can be arbitrary controlled by the observer. Therefore, a 
tritter, together with the 
three phase shift devices, performs the unitary transformation 
$\hat{U}(\vec{\phi})$ with the entries $U_{kl}=T_{kl}\exp(i\phi_l)$.

We calculate quantum probabilities in a standard way, i.e. 
\begin{eqnarray}
&&P_{QM}^{kl}(a_k;b_l)=
Tr(\Pi_{a_k}\otimes\Pi_{b_{l}}
\hat{U}(\vec{\phi}_k)\otimes\hat{U}(\vec{\theta}_l
|\psi\rangle\langle\psi|\hat{U}^{\dagger}(\vec{\phi}_k)\otimes\hat{U}^{\dagger}(\vec{\theta}_l))\nonumber\\
&&P_{QM}^{k}(a_k)=Tr(\Pi_{a_k}\otimes I|\psi\rangle\langle\psi|)\nonumber\\
&&Q_{QM}^{l}(b_l)=Tr(I\otimes\Pi_{b_l}|\psi\rangle\langle\psi|),
\end{eqnarray}
where $\vec{\phi}_k$ denotes the set of phase shifts at Alice's side 
when she measures
the observable $A_k$, $\vec{\theta}_l$ denotes the set of phase shifts 
at Bob's side when he measures the observable $B_l$ 
and $\Pi_{a_k},\Pi_{b_{l}}$ are projectors on the 
states $|a_k\rangle$, $|b_l\rangle$ respectively.

Following \cite{KASZLIKOWSKI}, we define the amount of violation of local
realism as the minimal noise admixture $F_{thr}$ to the state (\ref{state}) 
below which the measured correlations cannot be described by local
realism for the given observables. Therefore, we assume that Alice
and Bob perform their measurements on the following mixed state
$\rho_{F}$
\begin{eqnarray}
&&\rho_{F}=(1-F)|\psi\rangle\langle\psi|+F\rho_{noise},
\label{noise}
\end{eqnarray}
where $0 \leq F \leq 1$ and where $\rho_{noise}$ is a diagonal
matrix with entries equal to $1/9$. This matrix is a totally
chaotic mixture(noise), which admits a local
realistic description. For $F=0$ (pure maximally entangled state), a
local realistic description does not exist whereas for $F=1$ (pure
noise) it does. Therefore, there exists some threshold value of
$F$, which we denote by $F_{thr}$, such that for every $F\leq
F_{thr}$ local and realistic description does not exist. The bigger the value
of $F_{thr}$, the stronger is the violation of local realism. 
The quantum probabilities calculated (denoted with tilde)
on (\ref{noise}) acquire the form 
\begin{eqnarray}
&&\tilde{P}_{QM}^{kl}(a_k;b_l)=(1-F)P_{QM}^{kl}(a_k,b_l)+{F\over 9}\nonumber\\
&&\tilde{P}_{QM}^{k}(a_k)=P_{QM}^{k}(a_k)\nonumber\\
&&\tilde{Q}_{QM}^{l}(b_l)=Q_{QM}^{l}(b_l).
\end{eqnarray}

Let us now assume that Alice measures two observables defined by
the following sets of phase shifts
$\vec{\phi}_1=(0,\pi/3,-\pi/3)$ and $\vec{\phi}_2=(0, 0, 0)$, whereas Bob measures two
observables defined by the sets of phase shifts
$\vec{\theta}_1=(0, \pi/6, -\pi/6)$ and $\vec{\theta}_2=(0, -\pi/6, \pi/6)$. From numerical
computations, it is known \cite{KASZLIKOWSKI, DURT} that these sets
of phases gives the highest $F_{thr}$. Straightforward calculations
give the following values of the probabilities for each
experiment (please notice that we give here only probabilities for $F=0$):
\begin{eqnarray}
&&P_{QM}^{11}(1;1)=P_{QM}^{11}(2;3)=P_{QM}^{11}(3;2)={1\over 27}\nonumber\\
&&P_{QM}^{11}(1;2)=P_{QM}^{11}(2;1)=P_{QM}^{11}(3;3)={4+2\sqrt3\over 27}\nonumber\\
&&P_{QM}^{11}(2;2)=P_{QM}^{11}(1;3)=P_{QM}^{11}(3;1)={4-2\sqrt3\over 27}\nonumber\\
&&P_{QM}^{12}(1;1)=P_{QM}^{12}(2;3)=P_{QM}^{12}(3;2)={4-2\sqrt3\over 27}\nonumber\\
&&P_{QM}^{12}(1;2)=P_{QM}^{12}(2;1)=P_{QM}^{12}(3;3)={4+2\sqrt3\over 27}\nonumber\\
&&P_{QM}^{12}(2;2)=P_{QM}^{12}(1;3)=P_{QM}^{12}(3;1)={1\over 27}\nonumber\\
&&P_{QM}^{21}(1;1)=P_{QM}^{21}(2;3)=P_{QM}^{21}(3;2)={4+2\sqrt3\over 27}\nonumber\\
&&P_{QM}^{21}(1;2)=P_{QM}^{21}(2;1)=P_{QM}^{21}(3;3)={1\over 27}\nonumber\\
&&P_{QM}^{21}(2;2)=P_{QM}^{21}(1;3)=P_{QM}^{21}(3;1)={4-2\sqrt3\over 27}\nonumber\\
&&P_{QM}^{22}(1;1)=P_{QM}^{22}(2;3)=P_{QM}^{22}(3;2)={1\over 27}\nonumber\\
&&P_{QM}^{22}(1;2)=P_{QM}^{22}(2;1)=P_{QM}^{22}(3;3)={4+2\sqrt3\over 27}\nonumber\\
&&P_{QM}^{22}(2;2)=P_{QM}^{22}(1;3)=P_{QM}^{22}(3;1)={4-2\sqrt3\over 27}
\label{all}
\end{eqnarray}
whereas all the single probabilities are equal to ${1\over 3}$.
Putting (\ref{all}) into the inequality (\ref{inequality}) we find that it is not violated if
$F\leq {11-6\sqrt3\over 2}=F_{thr}$. This result is consistent 
with numerical result presented in \cite{KASZLIKOWSKI} as well as with the analytical proof
presented in \cite{CHEN}. Because, the proof showed in \cite{CHEN} 
gives neccesary and sufficient
conditions for the existence of local realistic description of the 
experiment considered, it strongly
suggests that the inequality (\ref{inequality}) is also a neccesary condition 
for a local realistic
description of quantum mechanical correlations observed for entangled pairs of qutrits.

\section{Appendix}
In this appendix we sketch the proof of inequality (\ref{inequality}).  
Let us consider the left hand side of the inequality (\ref{inequality}). 
It can be written as a sum of three parts
that we denote by $CH_1$, $CH_2$ and $G$ (we use the fact that probabilities 
appearing in the inequality 
can be written as marginals of the joint probability distribution)
\begin{eqnarray}
&&CH_1=\sum_{l=1}^{3}\sum_{n=1}^{3}P_{LR}(2,l;1,n)+
\sum_{l=1}^{3}\sum_{m=1}^{3}P_{LR}(2,l;m,1)
-\sum_{k=1}^{3}\sum_{n=1}^{3}P_{LR}(k,2;1,n)+
\sum_{k=1}^{3}\sum_{m=1}^{3}P_{LR}(k,2;m,1)\nonumber\\
&&-\sum_{l=1}^{3}\sum_{m=1}^{3}\sum_{n=1}^{3}P_{LR}(2,l;m,n)-
\sum_{k=1}^{3}\sum_{l=1}^{3}\sum_{m=1}^{3}P_{LR}(k,l;m,1),\nonumber\\
&&CH_2=\sum_{l=1}^{3}\sum_{n=1}^{3}P_{LR}(1,l;2,n)+
\sum_{l=1}^{3}\sum_{m=1}^{3}P_{LR}(1,l;m,2)
-\sum_{k=1}^{3}\sum_{n=1}^{3}P_{LR}(k,1;2,n)+
\sum_{k=1}^{3}\sum_{m=1}^{3}P_{LR}(k,1;m,2)\nonumber\\
&&-\sum_{l=1}^{3}\sum_{m=1}^{3}\sum_{n=1}^{3}P_{LR}(1,l;m,n)-
\sum_{k=1}^{3}\sum_{l=1}^{3}\sum_{m=1}^{3}P_{LR}(k,l;m,2),\nonumber\\
&&G=\sum_{l=1}^{3}\sum_{n=1}^{3}P_{LR}(2,l;2,n)+
\sum_{l=1}^{3}\sum_{m=1}^{3}P_{LR}(1,l;m,1)
-\sum_{k=1}^{3}\sum_{n=1}^{3}P_{LR}(k,2;2,n)+
\sum_{k=1}^{3}\sum_{l=1}^{3}P_{LR}(k,2;m,2).
\end{eqnarray}
Please notice that $CH_1$ and $CH_2$ are Clauser-Horne inequalitiess for pairs of detectors $1$ for Alice $2$ for Bob
and $2$ for Alice and $1$ for Bob respectively.
By summing every terms and rearranging if necessary, 
we get the following expression

\begin{eqnarray}
&&-(P_{LR}(1,1;1,1)+P_{LR}(1,1;1,3)
+P_{LR}(1,1;2,1)+P_{LR}(1,1;2,3)
+P_{LR}(1,1;3,1)+P_{LR}(1,1;3,3)\nonumber\\
&&+P_{LR}(1,2;1,1)+P_{LR}(1,2;1,2)
+2P_{LR}(1,2;1,3)+P_{LR}(1,2;2,3)
+P_{LR}(1,2;2,3)+P_{LR}(1,2;3,3)\nonumber\\
&&+P_{LR}(1,3;1,1)+P_{LR}(1,3;1,2)
+P_{LR}(1,3;1,3)+P_{LR}(1,3;3,1)
+P_{LR}(1,3;3,2)+P_{LR}(1,3;3,3)\nonumber\\
&&+P_{LR}(2,1;2,1)+P_{LR}(2,1;2,2)
+P_{LR}(2,1;2,3)+P_{LR}(2,1;3,1)
+P_{LR}(2,1;3,2)+P_{LR}(2,1;3,3)\nonumber\\
&&+P_{LR}(2,2;1,2)+P_{LR}(2,2;1,3)
+P_{LR}(2,2;2,2)+P_{LR}(2,2;2,3)
+P_{LR}(2,2;3,2)+P_{LR}(2,2;3,3)\nonumber\\
&&+P_{LR}(2,3;1,2)+P_{LR}(2,3;2,2)
+P_{LR}(2,3;3,1)+2P_{LR}(2,3;3,2)
+P_{LR}(2,3;3,3)+P_{LR}(3,1;1,1)\nonumber\\
&&+2P_{LR}(3,1;2,1)+P_{LR}(3,1;2,2)
+P_{LR}(3,1;2,3)+P_{LR}(3,1;3,1)
+P_{LR}(3,2;1,1)+P_{LR}(3,2;1,2)\nonumber\\
&&+P_{LR}(3,2;1,3)+P_{LR}(3,2;2,1)
+P_{LR}(3,2;2,2)+P_{LR}(3,2;2,3)
+P_{LR}(3,3;1,1)+P_{LR}(3,3;1,2)\nonumber\\
&&+P_{LR}(3,3;2,1)+P_{LR}(3,3;2,2)
+P_{LR}(3,3;3,1)+P_{LR}(3,3;3,2)),
\end{eqnarray}
which due to the positivity of the joint probability distribution 
$P_{LR}(a_1,a_2;b_1,b_2)$ is always negative or equal zero. This completes the proof.


\begin{references}


\bibitem{CH} J. F. Clauser and M. A. Horne, Phys. Rev. D {\bf 10}, 526 (1974).

\bibitem{FINE} A. Fine, Phys. Rev. Lett. {\bf 48}, 291 (1982).








\bibitem{KASZLIKOWSKI} D. Kaszlikowski, P. Gnaci\'nski, M. \.Zukowski, W. Miklaszewski and
A. Zeilinger, Phys. Rev. Lett. {\bf 85}, 4418 (2000).

\bibitem{DURT} T. Durt, D. Kaszlikowski and M. \.Zukowski,
quant-ph//0101084 (to be published in Phys. Rev. A. as brief report).


\bibitem{CHEN} Jing Ling Chen, Dagomir Kaszlikowski, L. C. Kwek, Marek \.Zukowski and C. H. Oh,
quant-ph//0103099.

\bibitem{CHSH} J. F. Clauser, M. A. Horne, A. Shimony and R. A. Holt,
Phys. Rev. Lett. {\bf 23}, 880 (1969).

\bibitem{TRITTERS} M. \.Zukowski, A. Zeilinger, M. A. Horne, Phys. Rev. A {\bf 55}, 
2564 (1997).


\end{references}
\end{document}